# Achievements, Experiences, and Lessons Learned from the European Research Infrastructure ERIGrid related to the Validation of Power and Energy Systems

[1]T. Strasser, [2]E.C.W. de Jong, [3]M. Sosnina, [4]J.E. Rodriguez-Seco, [5]P. Kotsampopoulos, [6]D. Babazadeh, [7]K. Mäki, [8]R. Bhandia, [9]R. Brandl, [10]C. Sandroni, [11]K. Heussen, [12]F. Coffele

[1]AIT Austrian Institute of Technology, [2]KEMA B.V., [3]DERlab e.V., [4]TECNALIA Research & Innovation, [5]National Technical University of Athens, [6]OFFIS e.V., [7]VTT Technical Research Center of Finland, [8]Delft University of Technology, [9]Fraunhofer Institute for Energy Economics and Energy System Technology, [10]Ricerca sul Sistema Energetico, [11]Technical University of Denmark, [12]University of Strathclyde

[1]Austria, [2,8]Netherlands, [3,6,9]Germany, [4]Spain, [5]Greece, [7]Finland, [10]Italy, [11]Denmark, [12]UK

## ABSTRACT

Power system operation is of vital importance and must be developed far beyond today's practice to meet future needs. Almost all European countries are facing an abrupt and very important increase of renewables with intrinsically varying yields which are difficult to predict. In addition, an increase of new types of electric loads and a reduction of traditional production from bulk generation can be observed as well. Hence, the level of complexity of system operation steadily increases. Because of these developments, the traditional power system is being transformed into a smart grid. Previous and ongoing research has tended to focus on how specific aspects of smart grids can be developed and validated, but until now there exists no integrated approach for analysing and evaluating complex smart grid configurations. To tackle these research and development needs, a pan-European research infrastructure is realized in the ERIGrid project that supports the technology development as well as the roll out of smart grid technologies and solutions. This paper provides an overview of the main results of ERIGrid which have been achieved during the last four years. Also, experiences and lessons learned are discussed and an outlook to future research needs is provided.

## KEYWORDS

Smart grids – simulation – hardware-in-the-loop – testing – research infrastructure – education – training

## 1   INTRODUCTION

Power system operation is of vital importance and has to be developed far beyond today's practice in order to meet future needs like the integration of Renewable Energy Resources (RES), Battery Energy Storage Systems (BESS), Electrical Vehicle Supply Equipment (EVSE) and other types of new devices and components [1], [2]. In fact, nearly all European countries are facing an abrupt and very important increase of RES such as wind and photovoltaic that are intrinsically variable with yields that are up to some extent difficult to predict. In addition, an increase of new types of electric loads such as air conditioning, heat pumps, and electric vehicles, and a reduction of traditional production from bulk generation can be observed. Hence, the level of complexity of system operation steadily increases. To avoid dramatic consequences, there is an urgent need for an extended system flexibility [3]. Also, the roll-out of smart grid technology, solutions, and corresponding applications based on recent developments in Information and Communication Technology (ICT) and power electronics is of particular importance in order to realize a number of advanced system functionalities (power/energy management, demand side management, ancillary services provision, etc.) [4], [5].

As a consequence of these developments, the traditional power system is being transformed into a Cyber-Physical Energy System (CPES), a so-called smart grid [1]-[5]. Previous and ongoing research have tended to focus on how specific aspects of smart grids can be developed and validated, but until now there exists no integrated approach for analysing and evaluating complex CPES configurations [6], [7].

In order to tackle the above identified research and development needs, a pan-European Research Infrastructure (RI) has been realized in the European ERIGrid project which supports the technology development as well as the roll out of smart grid technology and solutions [7]. It provides a holistic, CPES-based approach by integrating European research centres and institutions with outstanding power system and smart grid laboratory infrastructure to jointly develop common methods, concepts, and procedures and provide them for free to researchers, engineers, and students. The main goals of the ERIGrid project can be summarized as follows [7], [8]:

- The creation of a single point of reference supporting research and technology development on all aspects of smart grid systems validation,
- The development of a coordinated and integrated approach using the partners' expertise and RIs/laboratories more effectively, adding value to research and development projects,
- Facilitating a wider sharing of knowledge, tools, and techniques across fields and between academia and industry, and
- Accelerating pre-normative research and promoting the rapid transfer of research results into industrial-related standards to support future smart grids development, validation, and roll out.

The overall approach of ERIGrid is shown in Figure 1. Different Networking (NA), Joint Research (JRA), and Trans-national Access (TA) activities are covered by the project where the target is to realise the systematic validation and testing of smart grid configurations from a holistic, cyber-physical systems point of view. A multi-domain approach is realized, and corresponding methods and tools are developed which allow a systematic testing of smart grid applications covering power system and Information and Communication Technology (ICT) topics in an integrated manner. [7]

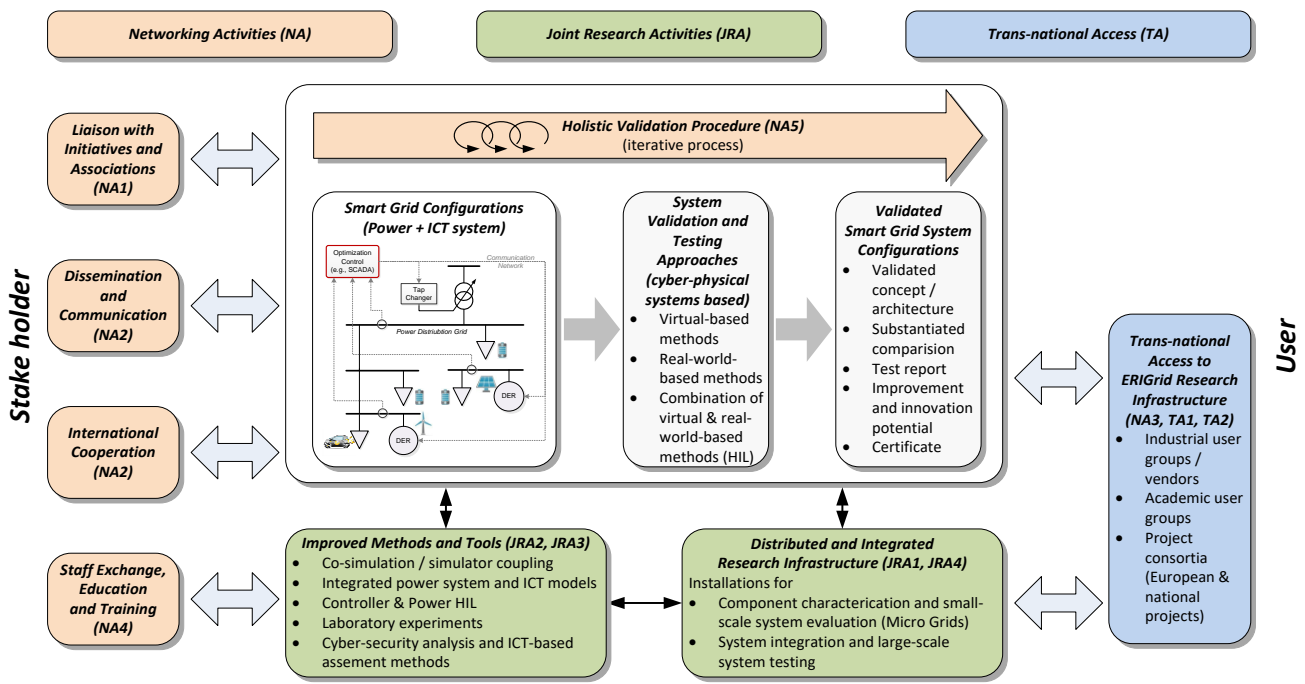

*Figure 1: Overview of the ERIGrid approach related to the validation and testing of smart grids [7]*

This paper provides an overview of the main achievements and results which have been carried out during the last four years in the ERIGrid project. Also, experiences and lessons learned are discussed. Therefore, Section 2 discusses the developed methods and tools for power system testing whereas Section 3 outlines the realized education and training concepts. The provided free access to the ERIGrid RIs/laboratories is discussed in Section 4. Finally, the paper concludes in Section 5 the main findings as well as an outlook about the necessary future research and development.

## 2 APPROACHES FOR POWER SYSTEM TESTING

### 2.1 Developed Methods and Tools

System-level validation of smart grid solutions can be a complex effort. A typical CPES solution, such as a distribution grid centralized demand response control system encompasses multiple disciplines (market, ICT, automation, infrastructure, etc.) and physical infrastructures (electricity, communication networks, etc.). Interactions among automation systems, enabling ICT, and electricity infrastructure are essential for such solutions and make testing of the integrated system a necessity [8].

As outlined above, ERIGrid tackled these issues by the provision of proper CPES-based validation concepts, methods, and corresponding tools. A procedural support can be useful when adopting a complex test platform attempting validation of a complex integrated control solution. A holistic view on testing procedures is illustrated in Figure 2(a) which is one of the core elements of the ERIGrid validation concept [8]-[10]. At the outset, this procedure template connects the system definition and use cases with a test objective in a test case. Once this link is fully established, the test specification fully captures the requirements for an experimental setup. The test platform can now be identified and suitably configured, even with a high level of complexity connecting several RIs/laboratories (e.g., RI *a* and RI *b*). The experiment execution in the corresponding RI/laboratory and the subsequent result evaluation may now lead to judging the test as successful, returning information with reference to the specifications and test case; or it may lead to a re-iteration of the specifications.

To document the testing needs and requirements as well as to specify the test case the corresponding experiment(s)templates have been developed in the project to support the whole validation chain. Figure 2(b) shows an example for such a template as a canvas. The description of the process and the corresponding templates can be downloaded from [10].

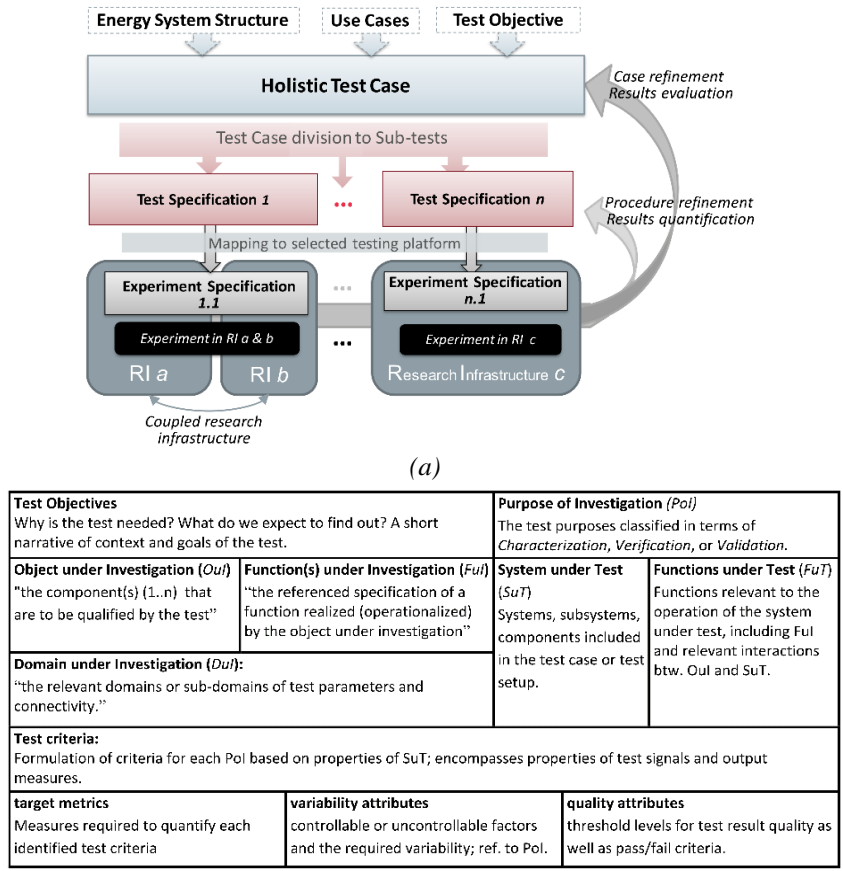

*(a)*

*(b)*

*Figure 2: Holistic test procedure: (a) overview of the approach, (b) test case template as canvas [8]*

Depending on the types of test purposes, relevant test platforms, devices or systems under test, etc., different procedures and methodologies are applicable. Under the conceptual frame of this holistic test procedure, the ERIGrid project defined specific approaches within co-simulation, multi-RI experiments, and hardware-in-the-loop testing. [8]

Figure 3 provides an overview of the proposed testing chain. In Stage 1 investigations performed in a pure software simulated environment are usually carried out in steady state or transient conditions. This enables the functionality test of the control algorithm but does not represent adequately the interface between power and control systems. Stage 2 of the testing chain proposes the use of two dedicated software tools for executing the power system model and controller separately. This Software-in-the-Loop (SIL) simulation or co-simulation technique allows the exchange of information in a closed loop configuration. After verifying the correct behaviour of the control algorithm in Stages 1 and 2, Stage 3 deals especially with the performance validation of the actual hardware controller using a Controller Hardware-in-the-Loop (CHIL) setup. CHIL testing provides significant benefits compared to simulation-only and SIL experiments. Using a Digital Real-Time Simulator (DRTS) for executing power system models in real time, the actual hardware controller can be tested including all kinds of communication interfaces and potential analogue signal measurements by interfacing it with the DRTS. [8]

The final Stage 4, before actual field-testing and implementation, of the proposed testing chain approach is the integration of real physical power hardware controlled by the hardware controller. This combined CHIL and Power Hardware-in-the-Loop (PHIL) is called Power System-in-the-Loop (PSIL) and includes the controller as well as power apparatus like inverter, motors, etc. This technique offers the closest possibility of a field test, which can be implemented in a laboratory, since it integrates real-time interactions between the hardware controller, the physical power component and the simulated power system test-case executed on the DRTS. Despite the high complexity to ensure stable, safe and accurate experiments, a PSIL setup enables an investigation, not as a single and separate entity, but as a holistic power system. This technique is proven to validate entire functionalities of real hardware controller, interdependencies and interactions between real power components in an entire flexible and repeatable laboratory environment. [8]

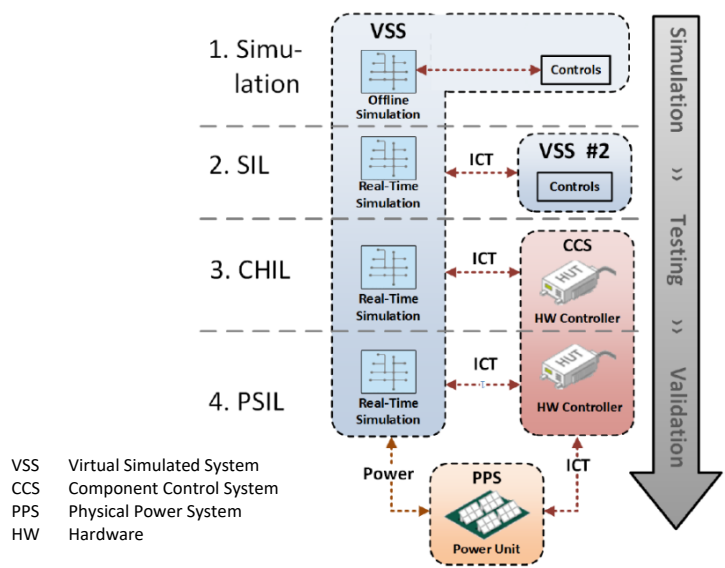

*Figure 3: Test chain concept for CPES validation [9]*

For executing experiments on different stages in the above outlined test chain concept ERIGrid improved and extended several concepts and methods. The main work on simulation-based tools was on the coupling of domain-specific simulators in a co-simulation manner. The Functional Mock-up Interface (FMI) approach has been used for interfacing different power system (PowerFactory, Matlab/SIMULINK, etc.) and ICT (ns3, etc.) simulators. A further activity was to combine real-time simulation and Hardware-in-the-Loop (HIL) for extended experimental possibilities. Also, HIL-based approaches have been improved (time-delay compensation, improved stability assessment, quasi static PHIL/PSIL, etc.).

For the provision of a Pan-European CPES/smart grid RI ERIGrid developed an approach which allows the online coupling of the different laboratories of the partners called JaNDER. This Multi-RI integration allows the realization of complex validation needs which probably cannot be implemented in a single RI/laboratory. JaNDER is a cloud platform for the exchange of information (measurements, control signals, laboratory asset descriptions) between geographically distributed laboratories by using a secure internet connection and therefore mainly concepts the automation and Supervisory Control and Data Acquisition Systems (SCADA) of the different RIs [8]. A sketch of the basic architecture is provided in Figure 4 which outlines the basis operational principles. Overall, three different data sharing services are provided by JaNDER [11]; i.e., *(i)* Level 1 (basis data sharing), *(ii)* Level 2 (IEC 61850-based communication, and *(iii)* Level 3 (CIM-based communication). For the common real time repository of shared data, a naming scheme for the different control commands and measurement signal has been provided as well.

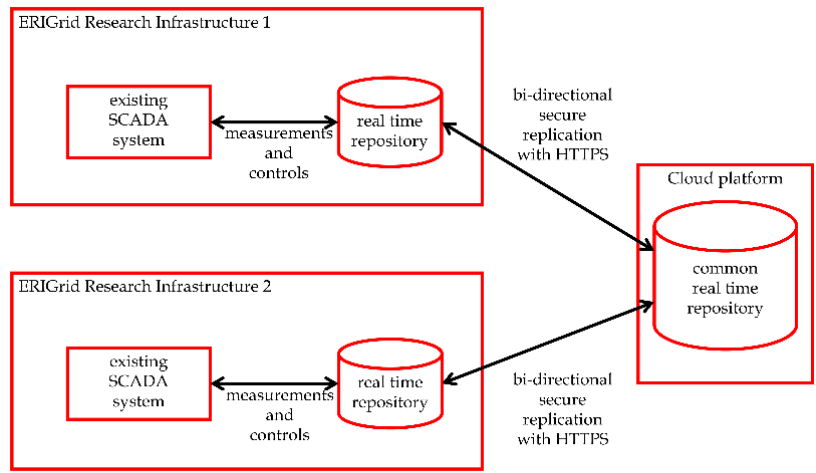

*Figure 4: Overview of the JaNDER architecture [8]*

Summarizing, most of the above outlined methods and tools are provided in an open access manner and can be accessed via the project website[1].

### 2.2 Experiences and Lessons Learned

The ERIGrid holistic validation approach has been broadly used for different project-internal validation activities, by several TA user groups to prepare and document the testing and validation work in the provided RIs/laboratories but also in a couple of other national and European projects.

Overall, it was highlighted by all the users of the holistic validation approach how the methodology promotes clarity and order when planning the experiments and helps also in the standardization of procedures between different RIs/laboratories, fostering the collaboration possibilities between different facilities. However, it turned out that there is complexity of the process and difficulty in understanding the involved concepts and definitions. The main steps in the future development of the approach must be oriented to enhance the clarity of the concepts and a simplification of the templates and documentation. Furthermore, an adoption of the validation approach to cover broader power and energy systems topics is also a future necessity to cope with future needs.

## 3 EDUCATION AND TRAINING OF POWER SYSTEM PROFESSIONALS

### 3.1 Overview of Developments

A need for new skills and expertise to foster the energy transition has risen, considering the increased complexity of CPES. Tackling the contemporary significant challenges requires a skilled workforce and researchers with systemic/holistic thinking and problem-solving skills. At the same time, technological advances can revolutionise education by allowing the use of new technical tools. [8]

---
[1] https://erigrid.eu/dissemination#open-access-tools

Due to the increased complexity of CPES, current and future engineers and researchers should have a broad understanding of topics of different domains, such as electric power, heat and definitely ICT related topics. Therefore, ERIGrid developed a comprehensive set of training material/activities for high school and university students, (young) researchers, and professionals. In this framework, e-learning tools and hands-on laboratory exercises dealing with important aspects of smart grids and distributed energy resources have been developed (selected examples are shown in Figure 5). These include the delivery of six webinars (on co-simulation, co-simulation with real-time simulation, hardware-in-the-loop simulation, information and communication technology standards for smart grids, holistic validation approach, multi RI/laboratory-based testing), the development of software tools (on co-simulation, interactive Jupyter notebooks, cyber-resilience tool, etc.), the organization of five training schools, and the creation of several presentations for teaching. [8]

The developed education material and corresponding tools are publicly available at the project website[2] to promote their use and replicability.

Summarizing, ERIGrid developed over 20 various educational resources, including software and programming tools, remote access to labs, lab exercises, webinars, course materials and other e-learning materials. Over 450 students have already applied ERIGrid exercises, tools, and other resources in their Master or PhD courses and theses. With over 15 educational events, including workshops, tutorials and schools, ERIGrid reached nearly 450 participants who highly appreciated the innovative lab sessions and the impact on their work and/or thesis. Also, around ERIGrid 300 real-time webinar participants have been reached with the above-mentioned webinars. Finally, ERIGrid granted to over 200 researchers, PhD students and young professionals, free access to world-class smart grids laboratories funded through the TA programme.

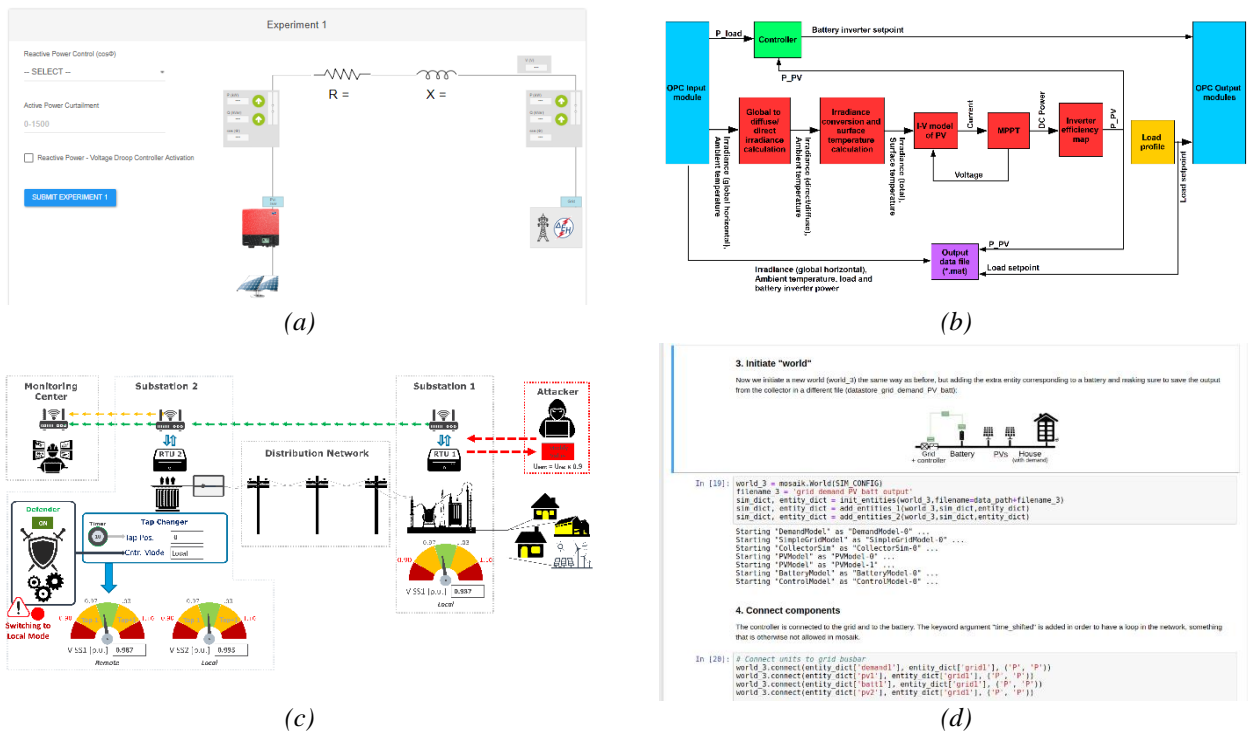

*Figure 5: Examples for ERIGrid education tools: (a) remote lab for voltage control experiments, (b) illustration of control blocks for microgrid remote lab balancing, (c) cyber-resilience tool, and (d) interactive Jupyter notebooks*

---

[2] https://erigrid.eu/education-training

*3.2 Experiences and Lessons Learned*

Most notably, the learners gain remote access to advanced laboratory infrastructures allowing to control equipment and monitor variables (i.e., remote labs). In addition, hands-on laboratory exercises are reported focusing on real-time simulation. Real-time power hardware-in-the-loop simulation was used for the first time for power engineering education on the important topic of distributed energy resource integration, which was appreciated by the students as shown from questionnaire statistics. Finally, educational activities targeting the younger generations are presented to disseminate the advantages of the smart grid and are able to increase the interest in pursuing a distributed energy resource engineering-related career path.

# 4 ACCESS TO EUROPEAN POWER SYSTEM AND SMART GRID LABORATORIES

*4.1 Overview of Implemented TA User Projects*

One of the core aims of the ERIGrid project is to provide system level validation methods and tools – which have been integrated into the partners' RIs/laboratories (i.e., 19 facilities spread over 11 European countries) – to external researchers, engineers and students. Overall, the project has realized six open calls for the so-called TA user projects where 97 submissions have been received. All those proposals have been carefully evaluated by around 55 independent experts from the domain and finally 75 projects (with more than 150 users) have been implemented in the RIs/laboratories of the partners. Therefore, around 1,000 access days (i.e., days of using the RIs/laboratories) have been provided by the ERIGrid partners to the implemented TA user projects. For the rest of the proposals, 8 have been rejected due to insufficient quality, 9 have been withdrawn by the corresponding proposers due to different reasons (persons not available anymore, lack of time, changed research agenda, etc.) and 5 more proposals could not be realized to due lack of available project budget.

Furthermore, the following list provides additional statistics/information about the supported TA user projects:

- 13 projects were led by companies,
- 19 projects had companies involved,
- 15 projects came from outside Europe,
- 8 projects were from ERIGrid partners ("internal TA"), and
- 4 multi-side projects (i.e., more than one ERIGrid RI/laboratory)

have been realized. The covered topics of all the implemented TA user projects had a wide range of activities which can be summarized as:

- Distributed energy resource and power system components characterization and evaluation,
- Testing of control concepts for power distribution operation,
- Microgrid controller testing,
- Analysis of the interaction of power and ICT systems in a smart grid context,
- Smart grid ICT/automation component validation, and
- Analysis of power electronic-based units covering the power and the controller part.

The outcomes of the above-mentioned TA user projects are publicly available at the project website[3].

*4.2 Experience and Lessons Learned*

Overall, the ERIGrid TA programme worked very well and supported quite a lot of different user groups in their research and development work which would have not been possible for them due to missing RIs/laboratory infrastructures in their institutions. The usage of the ERIGrid system-level validation concepts and corresponding tools was very helpful for the realization of the TA user projects. A good preparation of the experiments is crucial for an effective lab work during the access period, optimizing the available budget and lab time: it turned out that the holistic validation procedure was very helpful during the preparation phase of those projects (i.e., for the interaction and discussion of technical details between the TA user group and the hosting institution/laboratory).

---

[3] https://erigrid.eu/transnational-access

The good results have paved the way for the upcoming ERIGrid 2.0, where the same RIs/laboratories reinforced by three new installations have started the provision of access to external users. ERIGrid 2.0 will extend the developed testing and validation concepts and tools, with special focus on RI integration and automation (HIL and real-time simulation coupling, distributed co-simulation). Remote access to labs and virtual access to research resources (i.e., free of charge access to virtual data and services needed for research that are openly available through communication networks) will also be supported within ERIGrid 2.0.

# 5 CONCLUSIONS

The expected large-scale roll out of smart grid products and solutions during the next few years requires a multi-disciplinary understanding of several domains. The validation of such complex solutions gets more important than in the past. There is a clear shift from component-level to system-level testing. An integrated, cyber-physical systems-based, multi-domain approach for a holistic testing of smart grid solutions was missing, and is now specifically addressed by ERIGrid [7], [8].

Four research priorities have been successfully identified and tackled in this Pan-European project to overcome the shortcomings in today's validation and testing of power systems and corresponding components. The research focus was put onto the development of a holistic validation methodology and the improvement of simulation-based methods, hardware-in-the-loop approaches, and lab-based testing, which can be combined in a flexible manner. Furthermore, the integration and online connection of power systems/smart grid laboratories provides additional possibilities for the analysis and evaluation in ERIGrid. All these activities had to be supported by the development of training and education material for researchers and power system professionals [7], [8].

Summarizing, with all the achievements and results in the ERIGrid project powerful methods, concepts, and corresponding tools have been developed which allow a comprehensive testing of smart grid solutions. However, mainly the domains of power systems and ICT have been addressed by ERIGrid.

The provision of engineering and validation support will be critical for the successful development of future CPES applications and solutions. Without the proper tool support many of the tasks will require immense manual efforts and will require engineers educated in multiple domains (energy system physics, ICT, automation and control, cyber-security, etc.). The ERIGrid results provide a step forwards in the right direction, but more research efforts are still needed in the upcoming years to cope with new trends and corresponding requirements. Especially the integration of the electric power system with other domains (thermal, gas, water/waste water, transportation etc.) into a smart energy system requires additional efforts which are [12], [13]:

- Advanced RIs need to be developed which focus on the integration of different power and energy systems related areas (market issues, thermal topics, electric vehicles, etc.),
- A simplified access and corresponding services (facilitate future access by remote operation and coupling of both virtual and physical RI, etc.) to smart grid, smart energy systems, and renewables related RIs addressing challenging user needs is required,
- Domain-specific adaptations of previously developed abstract validation procedures and corresponding concepts, methods, and tools are required to address advanced applications (low-inertia grids, microgrids, hybrid grids, etc.),
- Common and well understood reference scenarios, use cases, and test case profiles for smart energy systems need to be provided to power and energy systems engineers and researchers; also, proper validation benchmark criteria and key performance indicators as well as interoperability measures for validating smart grids and smart energy systems need to be developed, extended, and publicly shared with domain professionals,
- A standardization of multi-domain CPES-based evaluation and testing procedures is necessary,

- Well-educated professionals, engineers, and researchers understanding smart grid and smart energy systems configurations in a multi-domain and cyber-physical manner addressing the upcoming energy transition need to be educated and trained on a broad scale.

The above listed open research and development issues are partly tackled by the successor project ERIGrid 2.0[4] which will be executed during the next years. It is planned to provide open access to numerous results of this project for all power and energy systems professionals.

## ACKNOWLEDGEMENT

This work is supported by the European Community's Horizon 2020 Program (H2020/2014-2020) under project "ERIGrid" (Grant Agreement No. 654113). Further information is available at the corresponding website www.erigrid.eu.

---

[4] https://erigrid2.eu